\documentclass[10pt,twocolumn]{article}
\usepackage[utf8]{inputenc}
\usepackage[T1]{fontenc}
\usepackage{lmodern}
\usepackage{amsmath,amssymb}
\usepackage{booktabs}
\usepackage{graphicx}
\usepackage[margin=0.72in]{geometry}
\usepackage{microtype}
\usepackage{xcolor}
\usepackage{hyperref}
\usepackage[numbers,sort&compress]{natbib}
\hypersetup{
  colorlinks=true,
  linkcolor=blue!50!black,
  citecolor=blue!50!black,
  urlcolor=blue!50!black,
  pdftitle={Uncertainty-Aware Airlight-Aligned Dual-Gain Recovery with Exact RGB Feasibility for Training-Free Single-Image Dehazing},
  pdfauthor={Maksim Sitnikov},
  pdfsubject={Training-free single-image dehazing},
  pdfkeywords={image dehazing, uncertainty, RGB feasibility, dual-gain recovery, total variation}
}
\newcommand{\method}{A$^2$CR}

\title{Uncertainty-Aware Airlight-Aligned Dual-Gain Recovery\\
with Exact RGB Feasibility for Training-Free Single-Image Dehazing}
\author{Maksim Sitnikov\\\small Independent Researcher}
\date{August 2026}

\begin{document}
\maketitle

\begin{abstract}
The inverse atmospheric-scattering model amplifies transmission, airlight, and sensor-noise
errors when a scalar gain $1/t$ is applied in dense haze. We introduce \method, a training-free
recovery operator that decomposes the airlight-centered RGB residual into components parallel
and perpendicular to the airlight vector. Two gains are obtained from analytic quadratic risks
containing transmission, airlight, and noise uncertainty. Each gain pair is constrained by the
exact convex polygon induced by the six RGB-cube inequalities and gain bounds. An edge-aware
joint total-variation problem is solved with a primal-dual algorithm and exact polygon projection.
On a controlled DIODE RGB-D evaluation comprising 500 frames and 22,500 reproducible haze
recipes, the full variant improves over scalar inversion by $2.951$ dB PSNR (95\% bootstrap CI
$[2.900,3.001]$), $0.191$ SSIM $[0.187,0.196]$, and $-0.978$ CIEDE2000
$[-1.074,-0.883]$, while eliminating post-recovery RGB violations. Results on four real paired
datasets are mixed and were not blind: the method is strongest on Dense-Haze but does not
dominate RFEP, DCP, and Haze-Lines elsewhere. We therefore claim an interpretable constrained
recovery study, not state-of-the-art dehazing. Code, 64 executable numerical and regression tests,
manifests, and statistical artifacts accompany the paper.
\end{abstract}

\section{Introduction}
Single-image dehazing is ill-posed: the observed linear-radiance color
\begin{equation}
 I(x)=J(x)t(x)+A(x)(1-t(x))
 \label{eq:asm}
\end{equation}
does not identify scene radiance $J$, transmission $t$, and airlight $A$ from one RGB sample.
Classical priors such as the dark channel prior (DCP) \cite{he2011dcp}, boundary constraints
\cite{meng2013boundary}, color attenuation \cite{zhu2015cap}, and haze-lines
\cite{berman2016nonlocal} make this ambiguity tractable without training. However, after any
estimate $\hat t,\hat A$ has been obtained, the common inverse
$J=\hat A+(I-\hat A)/\max(\hat t,t_{\min})$ applies the same gain to all color directions.
In dense haze, that gain amplifies both model error and chromatic sensor noise.

We study the recovery operator itself. The key observation is geometric: errors parallel to the
airlight vector and errors that change chromatic direction need not have the same reliability.
The proposed method applies separate analytic gains, but never permits them to leave the RGB cube.
Our contributions are:
\begin{itemize}
  \item an airlight-aligned two-component recovery with gains derived from explicit quadratic
        uncertainty risks;
  \item an exact per-pixel convex feasible polygon for the two gains, rather than post-hoc RGB
        clipping;
  \item a joint edge-aware TV formulation solved by primal-dual splitting with exact polygon
        projection;
  \item controlled and real-paired evaluations that report uncertainty, runtime, negative results,
        and the non-blind status of development datasets.
\end{itemize}

\section{Related Work}
\paragraph{Classical priors.}
DCP estimates haze thickness from dark pixels in local haze-free patches \cite{he2011dcp}.
Meng et al. constrain transmission through RGB boundaries and contextual regularization
\cite{meng2013boundary}. CAP estimates depth from HSV value and saturation \cite{zhu2015cap},
whereas non-local dehazing clusters clear colors into haze-lines in RGB space
\cite{berman2016nonlocal}. Bounded channel differences provide another deterministic color
constraint \cite{zhao2021bcdp}.

\paragraph{Spatial airlight and color.}
Airlight-field estimation relaxes the constant-airlight assumption \cite{zhang2018airlight}.
The color-constrained dehazing model jointly regularizes local airlight and transmission and uses
a learned natural-color distribution \cite{zhang2020color}. These works motivate spatial models,
but neither formulates two airlight-aligned recovery gains with an exact RGB-feasible set.

\paragraph{Uncertainty and learned dehazing.}
Semi-UFormer learns pixel uncertainty in a teacher-student model \cite{tong2022semi}, while BPUL
uses learned perturbation-induced uncertainty and degradation consistency \cite{wei2026bpul}.
Recent methods emphasize domain adaptation, diffusion, or unpaired transport
\cite{ma2025coa,wang2025hazing,lan2025dehazesb}. Our goal differs: no training, a deterministic
recovery operator, and an uncertainty model exposed in equations and diagnostic maps.

A targeted search of arXiv, CVF Open Access, and author publication pages through 1 August 2026
found no direct analogue combining airlight-aligned dual gains, analytic uncertainty risk, exact
RGB-feasible projection, and joint TV. This is a bounded literature-search statement, not proof
of worldwide priority.

\section{Method}
All physical operations use linear RGB. The final result alone is encoded to sRGB.

\subsection{Airlight-aligned decomposition}
For a global airlight estimate $A\in\mathbb{R}^3$, let $u=A/\|A\|_2$, $d=I-A$, and
\begin{equation}
 p=uu^\top d, \qquad q=(I_3-uu^\top)d.
\end{equation}
We recover
\begin{equation}
 J=A+g_{\parallel}p+g_{\perp}q.
 \label{eq:dual}
\end{equation}
The parallel component primarily changes distance from the airlight; the perpendicular component
changes chromatic direction.

\subsection{Analytic uncertainty risk}
For either component, with local signal energy $S$, mean transmission $\bar t$, transmission
variance $\sigma_t^2$, effective noise energy $N$, and identity prior $U$, we minimize
\begin{equation}
 R(g)=S[(g\bar t-1)^2+g^2\sigma_t^2]+Ng^2+U(1-g)^2.
 \label{eq:risk}
\end{equation}
The unconstrained minimizer is
\begin{equation}
 g^*=\frac{S\bar t+U}{S(\bar t^2+\sigma_t^2)+N+U},
 \quad 1\le g\le 1/t_{\min}.
 \label{eq:gain}
\end{equation}
Independent risks use component energies and projected airlight/noise uncertainty. When uncertainty
and noise vanish, Eq.~\eqref{eq:gain} reduces exactly to $1/t$; this limit is unit-tested.

Transmission uncertainty is estimated from an ensemble of DCP bootstrap, CAP, and a simplified
haze-line map. A weighted median and median absolute deviation are computed in optical depth
$D=-\log t$, then propagated as $\sigma_t^2\approx t^2\sigma_D^2$. Airlight uncertainty is the
component-wise variance of patch/top-percent bootstrap estimates.

\subsection{Exact RGB-feasible gain polygon}
For each pixel and channel $c$, Eq.~\eqref{eq:dual} must satisfy
\begin{equation}
 0\le A_c+g_{\parallel}p_c+g_{\perp}q_c\le1.
 \label{eq:rgb}
\end{equation}
Together with $1\le g_{\parallel},g_{\perp}\le g_{\max}$, the six half-planes in
Eq.~\eqref{eq:rgb} form a non-empty convex polygon because $(1,1)$ reproduces the valid input.
We enumerate feasible boundary intersections and edge projections in two dimensions and select the
Euclidean projection of the proposed gain pair. Randomized tests cover 20,000 direct projections
and 5,000 cached-polygon comparisons. No projected output violated the RGB cube.

\subsection{Joint edge-aware TV}
Let $K_x$ be the feasible polygon at pixel $x$. We solve
\begin{align}
 \min_{g_\parallel,g_\perp}\;&\sum_x R_{\parallel,x}(g_{\parallel,x})+
 R_{\perp,x}(g_{\perp,x})+\mu(g_{\parallel,x}-g_{\perp,x})^2 \nonumber\\
 &+\lambda\sum_x w_x\left(\|\nabla g_{\parallel,x}\|_2+
 \|\nabla g_{\perp,x}\|_2\right),
 \label{eq:joint}\\
 \text{s.t.}\quad &(g_{\parallel,x},g_{\perp,x})\in K_x\quad\forall x.\nonumber
\end{align}
where $w_x$ decreases across input-luminance edges. A Condat--Vu-style primal-dual splitting
algorithm \cite{condat2013} handles the smooth quadratic risk explicitly, TV through dual-ball
projection, and feasibility through the exact $K_x$ projection. Step sizes use a local Hessian
Lipschitz bound and $\|\nabla\|^2\le8$. The best feasible iterate by the original objective is
returned. A numerical unit test verifies objective decrease and feasibility at every pixel.

\subsection{Implementation and CUDA scope}
The reference implementation uses single-precision matrices for dense image operations and
double precision for scalar risk and projection calculations. A manually selected hybrid CUDA
backend moves the two full-resolution local-energy box filters to the GPU. The analytic risks,
exact polygon projector, and optional joint-TV iterations remain on the CPU. This preserves the
same objective and boundary convention; an executable regression compares final CPU and CUDA
outputs. We report this as an implementation path, not as a new algorithm or an end-to-end speedup.

\section{Experimental Protocol}
\paragraph{Controlled RGB-D synthesis.}
DIODE \cite{vasiljevic2019diode} supplies RGB, depth, and validity masks. We deterministically
select 250 indoor and 250 outdoor validation frames. Whole physical scene groups are assigned to
development/validation/internal-test splits (295/104/101 frames) without scene leakage. For each
frame we synthesize five target transmissions, three airlights, and three noise recipes in linear
RGB, yielding $500\times45=22{,}500$ recipes. Stable SHA-256-derived seeds make every streamed
sample reproducible without storing generated images.

We compare scalar inversion (B0), scalar RGB boundary projection (B3), dual gain with uncertainty
(B7), exact feasibility (B8), and joint TV (B9). Metrics include PSNR, SSIM
\cite{wang2004ssim}, CIEDE2000 \cite{sharma2005ciede}, chroma/hue error, RGB violations,
flat-region noise residual, and runtime. A 10-frame, 4,500-row pilot additionally computes LPIPS
AlexNet v0.1 \cite{zhang2018lpips}.

\paragraph{Uncertainty.}
The 45 recipes are first averaged within each frame. We then run 10,000 indoor/outdoor-stratified
paired bootstrap repetitions over 500 frames. Because DIODE validation contains only six physical
scene groups, we additionally report leave-one-scene-group-out ranges. Real-paired comparisons use
paired bootstrap by image identifier.

\paragraph{Real paired data.}
We use fixed test splits from I-HAZE \cite{ancuti2018ihaze}, O-HAZE \cite{ancuti2018ohaze},
Dense-Haze \cite{ancuti2019dense}, and NH-HAZE \cite{ancuti2020nh}. These datasets were inspected
during development; their results must not be called blind tests.

\section{Results}
\subsection{Controlled ablation}
Table~\ref{tab:controlled} summarizes 112,500 valid rows. B8 and B9 eliminate all RGB violations.
B9 has the best PSNR, SSIM, and flat-noise residual; B7 retains the best mean CIEDE2000 and chroma
error, so no universal color claim is supported.

\begin{table}[t]
\centering
\caption{DIODE controlled means over 22,500 recipes.}
\label{tab:controlled}
\resizebox{\columnwidth}{!}{%
\begin{tabular}{lrrrrrr}
\toprule
 & PSNR$\uparrow$ & SSIM$\uparrow$ & DE00$\downarrow$ & Chroma$\downarrow$ & Invalid$\downarrow$ & ms$\downarrow$\\
\midrule
B0 scalar & 18.797 & 0.515 & 12.521 & 12.672 & 0.259 & 0.30\\
B3 boundary & 19.965 & 0.611 & 13.758 & 16.556 & 0 & 0.35\\
B7 uncertain & 19.257 & 0.514 & \textbf{11.303} & \textbf{8.968} & 0.227 & 0.73\\
B8 feasible & 20.723 & 0.602 & 11.951 & 11.894 & 0 & \textbf{1.01}\\
B9 joint TV & \textbf{21.748} & \textbf{0.707} & 11.543 & 11.372 & 0 & 31.18\\
\bottomrule
\end{tabular}}
\end{table}

\begin{figure*}[t]
\centering
\includegraphics[width=0.96\textwidth]{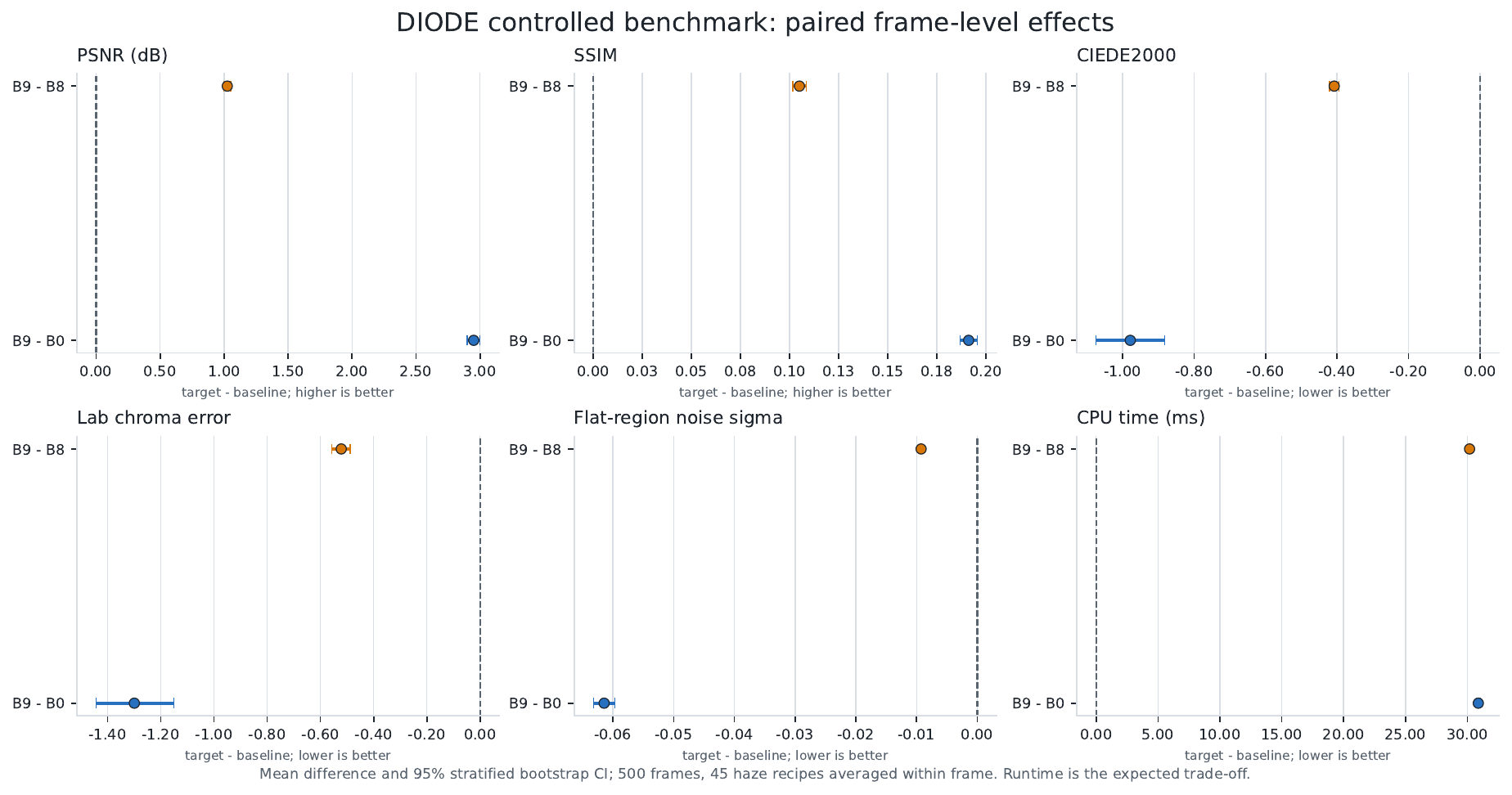}
\caption{Paired frame-level effects with 95\% stratified bootstrap intervals. Forty-five haze
recipes are averaged within each of 500 frames before resampling. Lower is better for CIEDE2000,
chroma error, noise, and runtime.}
\label{fig:bootstrap}
\end{figure*}

Against B0, B9 improves PSNR in 22,201/22,500 recipes, SSIM in all recipes, CIEDE2000 in 18,439,
and chroma error in 17,082. Figure~\ref{fig:bootstrap} shows frame-level intervals. All displayed
quality intervals exclude zero; the approximately 31-fold cost over B8 is the main trade-off.

The LPIPS pilot produced 4,500/4,500 finite values. B9 obtained PSNR 21.264, SSIM 0.595,
CIEDE2000 13.129, and LPIPS 0.624, compared with B0's 18.281, 0.358, 14.359, and 0.865.

\subsection{Real paired regression evaluation}
Table~\ref{tab:real} reports fixed defaults. A$^2$CR is best among the four tested methods on
Dense-Haze, but not elsewhere. On I-HAZE its LPIPS is significantly worse than RFEP in the paired
bootstrap; on O-HAZE SSIM improves while LPIPS worsens. NH-HAZE improvements depend on comparator
and metric. These negative results limit the claim to the recovery operator and controlled failure
modes.

\begin{table}[t]
\centering
\caption{A$^2$CR real-paired fixed test splits (not blind).}
\label{tab:real}
\resizebox{\columnwidth}{!}{%
\begin{tabular}{lrrrrr}
\toprule
Dataset & $n$ & PSNR$\uparrow$ & SSIM$\uparrow$ & DE00$\downarrow$ & LPIPS$\downarrow$\\
\midrule
I-HAZE & 15 & 16.616 & 0.822 & 13.659 & 0.280\\
O-HAZE & 22 & 17.046 & 0.798 & 14.342 & 0.331\\
Dense-Haze & 27 & 12.480 & 0.478 & 22.110 & 0.692\\
NH-HAZE & 27 & 13.228 & 0.602 & 20.901 & 0.462\\
\bottomrule
\end{tabular}}
\end{table}

\section{Limitations and Reproducibility}
The controlled benchmark depends on a synthetic scattering/noise family even though RGB and depth
are real. DIODE's 500 selected frames come from only six physical scene groups. The four paired
dehazing datasets were used during development. B9 is roughly 31 times slower than B8 at
maximum dimension 96. The hybrid CUDA path accelerates only local-energy filtering; the exact
projector and joint-TV solver remain CPU, and no end-to-end GPU speedup is claimed. B3 isolates a boundary
component and is not a complete reproduction of Meng et al. \cite{meng2013boundary}. No claim of
state-of-the-art quality or worldwide priority is made.

The repository records source URLs, archive hashes, frame manifests, stable seeds, commands,
raw CSV files, 10,000-bootstrap outputs, and 64 executable numerical and regression tests. It also distinguishes
raw primary metrics from optional aligned diagnostic metrics. Dataset licenses remain with their
authors; generated haze recipes are streamed and not redistributed.
Code and reproducibility artifacts are available at
\url{https://github.com/yellow444/SimpleDehaze}.

\section{Conclusion}
Airlight-aligned dual gains expose a useful degree of freedom hidden by scalar inverse scattering.
Analytic uncertainty risk, an exact RGB-feasible polygon, and joint TV produce strong controlled
improvements in structure, noise, and feasibility. Mixed real-paired results prevent a universal
color or perceptual claim. The next decisive experiment is a pre-registered blind external
evaluation, an oracle-$t/A$ recovery ablation, and comparison with complete external classical
baselines. The current hybrid CUDA implementation is useful for engineering evaluation but does
not alter the mathematical contribution.

\bibliographystyle{unsrtnat}
\bibliography{references}
\end{document}